\documentclass[10pt,conference]{IEEEtran}

\usepackage[hyphens]{url}
\usepackage{xspace}
\usepackage{siunitx}
\usepackage{amsmath}
\usepackage{amsfonts}
\usepackage{amssymb,bbm}
\usepackage{breqn}
\usepackage{mathrsfs} 
\usepackage{verbatim}
\usepackage{color}
\usepackage{graphicx,tikz}
\usepackage{wrapfig}
\usepackage{lipsum}

\def\un{\mathbbm{1}}
\def\;{\, ; \,}

\def\E{\mathbb{E}}

\def\SINR{\operatorname{SINR}}
\def\OP{\operatorname{OP}}
\def\OPAloha{\operatorname{OPAloha}}
\def\SNR{\operatorname{SNR}}

\def\ie{{\it i.e.},\xspace}

\title{2D Time-frequency interference modelling using stochastic geometry for performance evaluation in Low-Power Wide-Area Networks}
\author{\IEEEauthorblockN{        Zhuocheng
    Li\IEEEauthorrefmark{1}$^,$\IEEEauthorrefmark{2},       Steeve    Zozor\IEEEauthorrefmark{1},               Jean-Marc Brossier\IEEEauthorrefmark{1}, Nad\`ege Varsier\IEEEauthorrefmark{2} and Quentin
    Lampin\IEEEauthorrefmark{2}}
\IEEEauthorblockA{\IEEEauthorrefmark{1} 
Univ. Grenoble Alpes, GIPSA-Lab, F-38000 Grenoble, France\\
Emails: (zhuocheng.li,steeve.zozor,jean-marc.brossier)@gipsa-lab.grenoble-inp.fr}
\IEEEauthorblockA{\IEEEauthorrefmark{2} Orange Labs,
Meylan, France,
Emails: (zhuocheng1.li,quentin.lampin,nadege.varsier)@orange.com}}

\begin{document}
\maketitle


\begin{abstract}

  In wireless networks, interferences  between transmissions are modelled either
  in time or frequency domain. In this article, we jointly analyze interferences
  in the  time-frequency domain using  a stochastic geometry model  assuming the
  total time-frequency resources to be a two-dimensional plane and transmissions
  from  Internet  of  Things  (IoT)  devices  time-frequency  patterns  on  this
  plane.  To evaluate  the interference,  we  quantify the  overlap between  the
  information packets: provided that the  overlap is not too strong, the packets
  are not  necessarily lost due  to capture effect.  This flexible model  can be
  used for  multiple medium  access scenarios and  is especially adapted  to the
  random  time-frequency access  schemes  used in  Low-Power Wide-Area  Networks
  (LPWANs).  By  characterizing  the  outage  probability  and  throughput,  our
  approach  permits  to evaluate  the  performance  of  two representative  LPWA
  technologies        Sigfox{\textsuperscript        \textregistered}        and
  LoRaWAN{\textsuperscript \textregistered}.
\end{abstract}

%


\section{Introduction}
\label{Intro:sec}

Trading bit  rates for better link  budgets, LPWANs provide  long range wireless
connectivity    to    IoT   devices~\cite{centenaro2015long,    margelis2015low,
  vangelista2015long}.   Such  networks   provide  a  promising  alternative  to
traditional cellular or multi-hop networks  and are indeed envisioned to provide
nationwide connectivity  over industrial, scientific and medical  (ISM) bands to
battery-powered  IoT devices  that  transmit  little amount  of  data over  long
periods of  time, \emph{e.g.,} water \&  gas meters.  Thanks to  the long range,
the  IoT devices  can communicate  directly  with the  base stations  in a  star
topology.

Random  access schemes  such  as Aloha  are  commonly used  in  LPWANs in  which
multiple  devices access frequency  resources with  neither carrier  sensing nor
contention  mechanisms~\cite{Abr09, TanWet11}.   This reduces  the communication
overhead  and the  packet air  time,  but it  increases the  risk of  collisions
between  packets when  they overlap  in time  domain. Multiple  works  have been
dedicated to the interferences modelling in time domain~\cite{ware2001modelling,
  arnbak1987capacity,                                             cheun1998joint,
  lee2007experimental}.  In~\cite{cheun1998joint},  the  product  of  power  and
overlapping time duration between a  transmission of interest and an interfering
transmission is used  to represent the quantity of interfering,  then the sum is
taken over multiple interfering transmissions to give the total interference.

Note that interference modelling of transmission overlapping in frequency domain
is   also   well   studied   in   the  partially   overlapped   channels   (POC)
scenarios~\cite{mishra2006partially,  villegas2007effect,  xing2009multi}, which
are commonly used for networks such  as IEEE 802.11.  The interference factor in
the case of  POC is evaluated as the accumulated  energy in overlapped frequency
domain ~\cite{mishra2006partially}.  It has been  proven that the use of POC can
indeed  improve  the  network  throughput  in comparison  to  common  orthogonal
channelization schemes~\cite{xing2009multi, mishra2006partially}.

Our work  differs from the aforementioned  interference models in  that it's the
first, to the  best of our knowledge, to consider the  joint overlapping in both
time  and  frequency  domains. Our  model  based  on  stochastic geometry  is  a
high-level flexible one which can be adapted to multiple scenarios.

In section \ref{sec:Rework}, existing works on LPWANs performance evaluation are
introduced. In  section \ref{Model:sec}, our interference  modelling approach is
described and expressions of  $\SINR$, outage probability and network throughput
are given. Then in section \ref{Proba:sec}, we give the results on probabilistic
evaluation of overlapping. Finally in section \ref{App:sec}, the developed model
is   used  to   study   performances  of   two   different  LPWA   technologies,
Sigfox{\textsuperscript     \textregistered}     and    LoRaWAN{\textsuperscript
  \textregistered}.   Section  \ref{conclu:sec}   concludes   the  article   and
introduces some research perspectives.

\section{Related Work}
\label{sec:Rework}

Multiple   works    exist   for    the   performance   evaluation    of   LPWANs
\cite{mikhaylov2016analysis,   goursaud2016random,  adelantado2016understanding,
  reynders2016range}. Most  of these  works use Poisson  processes to  model the
packet arrival,  which we believe  is not the  most adapted for  periodic packet
sending scenarios in  LPWANs.  For example, a device reporting  on a daily basis
would not  send more than  one message per  day. However, as Poisson  models the
intensity of packet  arrival, an intensity of one message  per day represents in
fact the mean value, \ie one message  on average per day, which is not quite the
case    described.     In    \cite{mikhaylov2016analysis},    multiple    annuli
LoRaWAN{\textsuperscript  \textregistered} cell structure  is well  modelled and
illustrated with  a few applicative  scenarios. This structure is  considered in
our article but  channel effects and capture effect are added  to our model thus
making  it  more  complete  and realistic.   In  \cite{goursaud2016random},  the
performances  of  a random  Frequency  Division  Multiple  Access (random  FDMA)
scenario are studied in the pure  Aloha case, but the capture effect with little
overlap  between packets  is  not considered.  In \cite{reynders2016range},  the
performances   in   terms  of   packet   delivery   ratio   and  throughput   of
LoRaWAN{\textsuperscript     \textregistered}     and    Sigfox{\textsuperscript
  \textregistered}  are  simulated.  However,  the simulation  process  and  the
numerous network parameters are not  exposed enough thus lacking of transparency
and   possibility  of   reuse.    In  \cite{adelantado2016understanding},   some
interesting insights on  the limits of LoRaWAN{\textsuperscript \textregistered}
are given,  but again  the model  is based on  Poisson process  and there  is no
possible extension to account for the capture effect.  In this article, in order
to  give the  limit of  the  performance, we  study the  outage probability  and
throughput      of       LoRaWAN{\textsuperscript      \textregistered}      and
Sigfox{\textsuperscript  \textregistered}  when every  node  is transmitting  as
frequently as possible, according to  either the ISM band duty cycle constraints
or technology-related  constraints, which result  in message sending  periods in
the order of 1 to 10  minutes.  This scenario of the saturation throughput could
be that of packet  and object tracking systems~\cite{li2015internet}. Other less
frequent IoT scenarios such as water  \& gas metering can be evaluated using our
model thanks to its flexibility.


\section{The ``cards tossing'' model}
\label{Model:sec}

\subsection {Assumptions}

$N$ devices  $I_k$ share limited time-frequency  resources denoted by  $[0 \; T]
\times [0 \; F]$ to send packets to  a single base station, with $k = 0,1 \ldots
N-1$. $F$ is the bandwidth and $T$ the message sending period. $I_0$ is the user
of interest.  $g_k(t,f)$ the packet  sent by user  $I_k$. We make  the following
assumptions:
\begin{enumerate}
\item\label{limited:as}   Messages  from   all   the  senders   have  the   same
  rectangle-shaped  time-frequency  support,  since  they are  limited  in  time
  duration, denoted  by $\Delta t$  and bandwidth occupancy, denoted  by $\Delta
  f$.  Then $g_k$  takes the form $g_k(t,f) =  m_k(t,f) \, \un_{I_k}(t,f)$ where
  $\un_A$ stands for the indicator function of  set $A$ and $I_k = [t_k \; t_k +
  \Delta t] \times  [f_k \; f_k + \Delta f]$  denotes the time-frequency support
  of node $I_k$ (for the sake of  simplicity, we use the same notation $I_k$ for
  the sender itself and the time-frequency support of its message); $t_k$ is the
  initial time  of transmission  and $f_k$ the  lowest frequency of  the packet.
  $t_k  \in \left[0;T  -\Delta  t \right]$,  and  $f_k \in  \left[0;F -\Delta  f
  \right]$ (see figure~\ref{Aloha:fig}). 
\item\label{independent:as}  There  is   no  cooperation  between  the  devices,
  \emph{i.e.,}  random  access  considered.  The couples  $(t_k,f_k)$  are  thus
  independent.   They are also  assumed to  be uniformly  distributed, as  it is
  probably optimal in terms of dispersing the packets and avoiding collisions.
\item\label{uniform:as} Given  support $I_k$,  the time-frequency energy  of the
  information   packet  is  uniformly   distributed  over   $I_k$,  \emph{i.e.,}
  $\E\left[|g_k(t,f)|^2  \,  | \,  I_k\right]  =  \rho_k \un_{I_k}(t,f)$,  where
  $\rho_k$ is the energy density.
\item\label{noise:as}  The channel  is  affected by  an additive  time-frequency
  white   noise   $\xi(t,f)$    of   energy   density   $\gamma$,   \emph{i.e.,}
  $\E\left[|\xi(t,f)|^2\right] = \gamma$.
\end{enumerate}

Assumption~\ref{uniform:as}  is  the  ideal  and  most efficient  way  of  using
time-frequency   resources~\cite{mishra2006partially}.  In   practice,  transmit
spectrum mask is usually applied to specify the upper limit of power permissible
and  attenuate the signal  outside the  mask. An  attenuation of  $30\si{dB}$ to
$50\si{dB}$ is  observed in real-world  scenarios~\cite{mishra2006partially}, so
this assumption  can be  considered as realistic.  Notice that the  common Aloha
scenario is  encompassed in this  formalism by fixing  $\Delta f = F$.   In this
case, $f_k = 0$ but $t_k$ remains random.


\subsection{$\SINR$ expression }
\label{QuantitiesInterest:sec}

As mentioned  in the introduction,  interference can be  modelled as the  sum of
accumulated energy in  time-frequency domain, which is calculated  as the sum of
energy coming from different  interfering transmissions.  The $\SINR$ is defined
as  the ratio  between the  energy of  the message  of  interest, $\displaystyle
\int_{I_0}  \E\left[|g_0(t,f)|^2\right] \,  \mathrm{d}t  \,\mathrm{d}f$ and  the
interference    $\displaystyle     \sum_{k=1}^{N-1}    \int_{I_0    \cap    I_k}
\E\left[|g_k(t,f)|^2\right]   \,    \mathrm{d}t   \,\mathrm{d}f$   plus   noise,
\emph{i.e.,}

\begin{equation}
\SINR = \frac{\rho_0 \, \Delta t \, \Delta f}{\sum_{k=1}^{N-1}\rho_k \, S_{k} +
\gamma \, \Delta t \, \Delta f}
\end{equation}
where  $S_k  = \mu\left(  I_0  \cap  I_k \right)$  is  the  surface between  the
transmission  of interest  $I_0$  and an  interfering  one $I_k$  ($\mu$ is  the
surface measure).  We can  normalize $S_k$  with respect to  the surface  of the
time-frequency  support of  transmissions,  \ie $X_k  =  \frac{S_k}{\Delta t  \,
  \Delta f}$. Thus, the $\SINR$ can be recast as
\begin{equation}
\SINR = \frac{\rho_0 }{\sum_{k=1}^{N-1}\rho_k \, X_{k} + \gamma}
\label{SINR_XS:eq}
\end{equation}

\begin{figure}[ht]
\centerline{\includegraphics[width=.85\columnwidth]{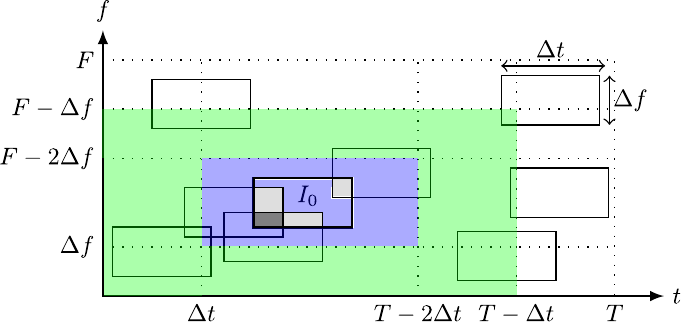}}
\caption{Illustration  of  the  ``cards  tossing'' game,  where  each  rectangle
  represents the information packet $I_k$.  The rectangle in bold represents the
  transmission of interest, \emph{i.e.,} the  packet $I_0$, while the gray areas
  depict  the sub  regions in  collision (the  darker the  area, the  larger the
  number of ``cards''  covering the sub region). The light  blue area is defined
  as the non border  area $\left[\Delta t;T-2\Delta t\right] \times \left[\Delta
    f;F-2\Delta f\right]$, denoted by $\overline{B}$. The light green defined as
  the  border area $\left[0;T-\Delta  t\right] \times  \left[0;F-\Delta f\right]
  \backslash  \overline{B}$.   These  two  areas   constitute  $\left[0;T-\Delta
    t\right] \times \left[0;F-\Delta f\right]$.}
\label{Aloha:fig}
\end{figure}

The overlapping  phenomenon between  packets is similar  to the game  of players
tossing cards  onto a table and  trying to recognize their  own cards afterwards
(see figure \ref{Aloha:fig}).  When there are too many  players, the probability
of  overlapping will  increase to  the extent  that it's  highly probable  to be
unable  to recognize  a  card. In  the  next subsection,  we  illustrate in  two
different  scenarios,  the  interest  of  our ``cards  tossing''  model  in  the
derivation  of the  outage probability  and  throughput of  wireless systems  in
function of the number of devices $N$.

\subsection{Outage and throughput model}
\subsubsection{With multipath fading and path loss}
\label{sec:outage1}
The first  scenario is when the packets  from devices suffer from  path loss and
fading. $\rho_k$ can be thus expressed as $\rho_k = \rho_{tm} h\,l(r_k)$.

\begin{itemize}
\item $\rho_{tm}$ models  the transmission energy density, which  is supposed to
  be identical for all devices.
\item  $l(r_k)$  models the  distance-dependent  attenuation, \emph{e.g.,}  path
  loss, where  $r_k$ is the Euclidean  distance between device $k$  and the base
  station.     We     choose     the     following    non     singular     model
  \cite{aljuaid2010investigating} expressed as  $l(r_k) = \alpha \left[\max(r_k,
    r_c)\right]^{-\beta}$, where $r_c$ is  a critical distance to avoid $l(r_k)$
  taking infinity when $r_k$ tends to 0.  Here we fix it to $1 \si{m}$. $\alpha$
  is a constant  modelling system-level losses and gains which is  fixed to 1 in
  our study.  $\beta$ is the path loss exponent assumed to be greater than 2.
\item $h$ is  a random variable modelling small  scale, large-scale or composite
  distance  non dependent  fading.  We  suppose that  $\sqrt{h}$ results  from a
  rayleigh multipath fading which  gives $h$ exponential cumulative distribution
  function (cdf), \ie $P_H(h) = 1 - \exp(-\lambda h)$ (with $P$ denoting the cdf
  and  $\E(h) =  \lambda$).  The  mean value  $\lambda$  is fixed  to  1 in  our
  study. Note that other forms of fading can be considered with our paradigm.
\end{itemize}

The $\SINR$ can be thus recast as follows,
\begin{equation}
\SINR = \frac{h}{\frac{\sum_{k=1}^{N-1}r_k^{-\beta} \, h \, X_{k}}{r_0^{-\beta}}
+ \frac{\gamma}{\rho_{tm} \,r_0^{-\beta}}}
\end{equation}

In order  to study the distance distribution  of devices in the  cell, we define
$r_{\max}$ as the  distance to the base station of the  most distant devices, in
the sense that their transmissions barely satisfy the target $\SINR$, denoted by
$\zeta$   in   the   presence   of   only  path   loss,   \ie   $\frac{\rho_{tm}
  r_{\max}^{-\beta}}{\gamma}   =    \zeta$.   This   gives    us   $r_{\max}   =
\sqrt[\beta]{\frac{\rho_{tm}}{\zeta \gamma}}$.

Let  us denote  $\Pr$ the  probability measure.   Suppose that  every  packet is
repeated  $n_{\mathrm{rep}}$ times,  and  the repetitions  are independent.  The
outage    probability    can    be    defined   as    $\left[\Pr\left[\SINR    <
    \zeta\right]\right]^{n_{\mathrm{rep}}}$,  which   is  further  expressed  as
follows,

\begin{equation}
\OP_{n_{\mathrm{rep}}}(r_0) = \left[\Pr\left[h < \frac{ r_{\max} ^{-\beta}}{
r_0^{-\beta}} + \frac{\zeta \sum_{k=1}^{N-1}r_k^{-\beta} \, h \,
X_{k}}{r_0^{-\beta}}\right]\right]^{n_{\mathrm{rep}}}
\label{eq:OPr0rep}
\end{equation}

Note that  repetition mechanism  is a  commonly used scheme  in LPWANs  to trade
efficiency for robustness of transmission.

Naturally, $\OP_{n_{\mathrm{rep}}}(r_0) $ depends  on the position of the device
of   interest.  Distant   devices  with   larger  $r_0$   suffer   from  greater
$\OP_{n_{\mathrm{rep}}}(r_0)$.   Suppose   that   the  devices   are   uniformly
distributed in the cell of the base station, which is defined as in the shape of
an annulus formed  with smaller radius $r_c$ and  larger radius $r_{\max}$.  The
probability  density function  (pdf)  of $r_k$  can  be expressed  as $p_R(r)  =
\frac{2r}{r_{\max}^2 - r_c^2}$ ($p$ is used to denote the pdf). The notation $k$
is omitted for the sake of simplicity.

Global outage probability  $\overline{\OP}_{n_{\mathrm{rep}}}$ is defined as the
outage probability averaged over $r_0$, \emph{i.e.,}
\begin{equation}
\overline{\OP}_{n_{\mathrm{rep}}} = \int_{r_c}^{r_{\max}}
\OP_{n_{\mathrm{rep}}}(r_0)\, \frac{2r_0}{r_{\max}^2 - r_c^2} dr_0
\label{eq:OPavg}
\end{equation}

The  effective throughput is  defined as  the average  number of  non repetitive
packets  received per unit  time and  is denote  by $\it{Th}(n_{\mathrm{rep}})$,
which can be expressed as follows,
\begin{equation}
\it{Th}(n_{\mathrm{rep}}) = \frac{N(1 - \overline{\OP}_{n_{\mathrm{rep}}})}{T \, n_{\mathrm{rep}}}
\label{eq:Throughput}
\end{equation} 
Recall that $N$ is the number of devices, and $T$ the message sending period. In
the case of pure Aloha, \ie packets considered lost when they collide in time or
frequency  domain, \ie  $X_{\Sigma} =  \sum_{k=1}^{N-1} X_k  \neq 0  $. Assuming
$X_k$ and $r_k$ independent, our outage probability can be recast as,

\begin{multline}
 \OPAloha_{n_{\mathrm{rep}}}(r_0) = \\ \left[\Pr\left[ X_{\Sigma} \neq 0 \right]
+ \Pr\left[ X_{\Sigma} = 0 \right] \Pr\left[ h < \frac{
r_\max^{-\beta}}{r_0^{-\beta}} \right] \right]^{n_{\mathrm{rep}}}
\label{eq:OPAloha}
\end{multline}

One can  observe that \eqref{eq:OPAloha} is greater  than \eqref{eq:OPr0rep}, as
\eqref{eq:OPr0rep}  includes  the  capture   effect,  \ie  certain  packets  not
considered       lost        even       in       case        of       collision.
$\overline{\OPAloha}_{n_{\mathrm{rep}}}$   and  $\it{ThAloha}(n_{\mathrm{rep}})$
can     be    calculated     in    the     similar    way.     By    definition,
$\overline{\OPAloha}_{n_{\mathrm{rep}}}  \leq \overline{\OP}_{n_{\mathrm{rep}}}$
and $\it{ThAloha}(n_{\mathrm{rep}}) \leq \it{Th}(n_{\mathrm{rep}})$.

\subsubsection{With perfect power control}
\label{sec:powerC}
We  consider another  scenario in  which the  packets of  different  devices are
supposed  to arrive  at  the base  station  with identical  energy density,  \ie
$\rho_0 =  \rho_k =  \rho$, thanks  to a certain  power control  mechanism.  The
$\SINR$ can be recast in this case as,
\begin{equation}
\SINR = \frac{1}{X_{\Sigma}  + \SNR^{-1}}
\label{eq:SINRTPC}
\end{equation} 
where $ \SNR = \frac{\rho}{\gamma}$. The outage probability  can be recast as,
\begin{equation}
\OP_{n_{\mathrm{rep}}} = \left[\Pr\left[X_{\Sigma} \ge \zeta^{-1} - \SNR^{-1}
\right]\right]^{n_{\mathrm{rep}}}
\label{eq:OP1D}
\end{equation} 

In this case the  non fairness between devices in terms of  distance to the base
station  is resolved.   $\OP_{n_{\mathrm{rep}}}$ does  not depend  on  $r_0$ any
more, but  only on $\gamma$, the  target $\SINR$ $\zeta$ and  $\rho$, the energy
density  that results from  the power  control.  The  average throughput  can be
recast as $\it{Th}(n_{\mathrm{rep}}) = \frac{N(1 - \OP_{n_{\mathrm{rep}}})}{T \,
  n_{\mathrm{rep}}}$.   The   quantities  to   be   simulated   are  listed   in
table~\ref{tab:simulQ}.

In  both scenarios,  we should  first study  the probabilistic  distributions of
$X_k$  and  $X_{\Sigma}$. In  the  next  section,  we derive  the  probabilistic
evaluations of $X_k$ and $X_{\Sigma}$. In  the case of multipath fading and path
loss,  the exact distribution  of $\sum_{k=1}^{N-1}r_k^{-\beta}  \, h  \, X_{k}$
remains difficult  to evaluate even with  the distribution of  $X_k$ derived and
$r_k$, $h$ and  $X_k$ assumed to be independent random  variables.  We use Monte
Carlo method to evaluate it.

\section{Probabilistic Evaluations}
\label{Proba:sec}

The results of $X_k$ and $X_{\Sigma}$ are  different in 1D case and 2D case.  We
first give the  results of $X_k$ in the easier 1D  case in section \ref{sec:1D},
\ie $\Delta f  = F$ so interference  happens only when there is  overlap in time
domain. Physical  layer technologies such  as spreading spectrum fall  into this
case.  Then the  results of $X_k$ in  the more complicated 2D case  are given in
section \ref{sec:2D},  where overlapping can  happen in both time  and frequency
domains, random FDMA approach belongs to this case.  The results on $X_{\Sigma}$
are given in section \ref{sec:sigma}.

We   denote   by  $p_{X_k}$   (resp.   $p_{X_{\Sigma}}$)   the   pdf  of   $X_k$
(resp.$X_{\Sigma}$),  by  $P_{X_k}$   (resp.  $P_{X_{\Sigma}}$)  the  cumulative
distribution function (cdf) of $X_k$ (resp. $X_{\Sigma}$).

\begin{figure}[ht]
\centerline{\includegraphics[width=0.85\columnwidth]{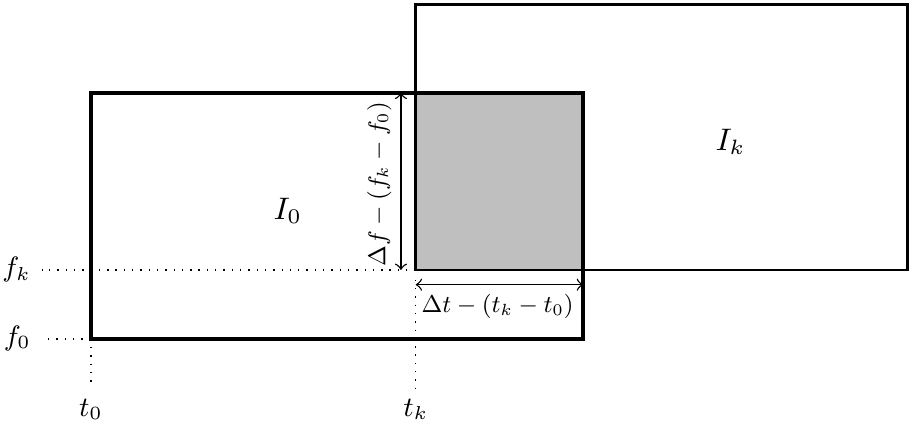}}
\vspace{-1mm}
\caption{When an  emission $I_k$ collides  with $I_0$, the overlapped surface,
  represented in gray, is $(\Delta t - |t_k-t_0|) (\Delta f - |f_k-f_0|)$.}
\label{Surface2:fig}
\end{figure}

The overlapped surface between two packets $X_k$ is determined by their relative
position in  $\left[0;T\right] \times \left[0;F\right]$.  Recall  that $t_k$ and
$f_k$ are  defined over$\left[ 0  \; T-  \Delta t \right]$  and $\left[ 0  \; F-
  \Delta f  \right]$. We can  thus define $\tau_k =  \frac{|t_k-t_0|}{\Delta t}$
and $\varphi_k = \frac{|f_k-f_0|}{\Delta f}$ as the normalized absolute time and
frequency   difference   between   emission   $I_0$  and   $I_k$,   see   figure
\ref{Surface2:fig}. $\tau_k$ and $\varphi_k$  are defined over $\left[ 0 \;N_t-1
\right]$  and $\left[  0 \;  N_f-1  \right]$ respectively.  From the  assumption
\ref{independent:as},      $(\tau_k,\varphi_k)$     also      have     identical
distributions.  For  the  sake  of  brevity,  we will  omit  index  $k$  in  the
expressions,  \emph{i.e.,}  pdf  of  $(\tau_k,\varphi_k)$  (resp.  $\tau_k$  and
$\varphi_k$)   is   denoted   by   $p_{\tau,\varphi}$  (resp.   $p_{\tau}$   and
$p_{\varphi}$) ,  and the (cdf) denoted by  $P_{\tau,\varphi}$ (resp. $P_{\tau}$
and $P_{\varphi}$).

\subsection{1D ``cards tossing'' game}
\label{sec:1D}

In the 1D game, $f_0 = f_k =  0$ so that $\varphi_k = 0$.  $f_k$ and $\varphi_k$
become deterministic and independent of $t_k$. We have
\begin{equation}
X_k = (1-\tau_k) \un_{[0 \; 1)}(\tau_k)
\end{equation}

One can easily  deduce that
\begin{equation}
\Pr[X_k > x] = \int_0^{1-x} p_{\tau}(u) du 
\end{equation}
which gives $\Pr[X_k > x] = P_{\tau}(1-x)$ and the probability of collision $p_c
= \Pr[X_k > 0]$ is given by $P_{\tau}(1)$.

In the case  where $t_k$ are assumed uniformly distributed over  $\left[ 0 \; T-
  \Delta t \right]$, the cdf of $X_k$, $P_{X_k}$ can be derived as follows,
\begin{equation}
P_{X_k} = 1 - \Pr[X_k > x] = 1 - \frac{(2 N_t-3 + x) (1-x)}{(N_t-1)^2}
\end{equation}
where $x \in [0  \; 1)$. $p_{X_k}$ can be obtained by  deriving $P_{X_k}$ for $x
\ne 0$, and $p_{X_k} (0) = P_{X_k} (0) = 1 - \frac{2 N_t-3}{(N_t-1)^2}$.

\subsection{2D ``cards tossing'' game}
\label{sec:2D}

Recall that  $t_k$ and $f_k$  are assumed independent and  uniformly distributed
over  $\left[  0  \;  T-  \Delta  t  \right]$ and  $\left[  0  \;  F-  \Delta  f
\right]$.  Denote $\frac{T}{\Delta  t}$  by $N_t$  and  $\frac{F}{\Delta f}$  by
$N_f$.

A simple look at the geometrical configuration plotted in figure~\ref{Surface2:fig}
allows to express the normalized surface as
\begin{equation}
X_k = (1-\tau_k) (1-\varphi_k) \un_{[0 \; 1)^2}(\tau_k,\varphi_k)
\end{equation}
Thus,  for  $x  \in  [0  \;  1)$,  we  have  $\Pr[X_k  >  x]  =  \Pr[(1-\tau_k)
(1-\varphi_k) >  x]$, we immediately
get,
\begin{equation}
\Pr[X_k > x] = \int_0^{1-x} \! \left( \int_0^{1-\frac{x}{1-u}}
\!\! p_{\tau,\varphi}(u,v) \, dv \right) du
\label{N2_ccdf:eq}
\end{equation}
where the bound of the first integral is  due to the fact that when $u \ge 1-x$,
$1-\frac{x}{1-u}  \le  0$,  the  inner  integral is  zero.  The  probability  of
collision  is  $p_c  =  P_{\tau,\varphi}(1,1)$  .   When  $t_k$  and  $f_k$  are
independent and uniformly distributed over $\left[ 0 \; T- \Delta t \right]$ and
$\left[ 0 \; F- \Delta f \right]$, some long algebra leads to,
\begin{equation}\begin{array}{l}
\displaystyle P_{X_k}  = 1 - \frac{(a + b \, x) (1-x) + (c + x) x \ln
x}{(N_t-1)^2 (N_f-1)^2}
\\[2mm] \mbox{with} \\[2mm]
\left\{\begin{array}{lll}
a & = & (2 N_t-3)(2 N_f-3)\\[2mm]
b & = & 9 - 2 N_t - 2 N_f\\[2mm]
c & = & 2 \, (N_t-2) (N_f-2)
\end{array}\right.
\end{array}\end{equation}
Similar procedures as in 1D game should be taken to find $p_{X_k}$.

\subsection{Probabilistic evaluation of $X_{\Sigma}$}
\label{sec:sigma}

Let's  first consider  the probabilistic  evaluation of  $X_{\Sigma}$ in  the 2D
case. Notice  that $\left[0;T-\Delta t\right]  \times \left[0;F-\Delta f\right]$
can  be divided  into two  areas  \emph{i.e.,} the  non border  area denoted  by
$\overline{B}$  and the  border area  denoted by  $B$. We  have  $\overline{B} =
\left[\Delta t;T-2\Delta t\right] \times \left[\Delta f;F-2\Delta f\right]$, and
$B   =   \left[0;T-\Delta    t\right]   \times   \left[0;F-\Delta   f\right]   /
\overline{B}$. See  figure \ref{Aloha:fig}. For  the sake of brevity,  we define
the event that  $(t_0, f_0)$ falls into the non  border area $\overline{B}$ also
as $\overline{B}$.  The event that  $(t_0, f_0)$ falls  into the border  area as
$B$.

In fact  $p_{X_k \vert \overline{B}}  \ne p_{X_k \vert  B}$ because a  packet in
$\overline{B}$ has  greater chance to be  corrupted by an interfering  one as it
can come from all directions. A  packet in $B$ cannot be interfered from certain
positions of $(t_k,  f_k)$ out of border, resulting in  a smaller probability of
being  corrupted.  In  section  \ref{sec:1D}  and \ref{sec:2D},  we  could  have
separated  the  derivation in  $\overline{B}$  and  $B$  and obtained  the  same
results.  For   $P_{X_{\Sigma}}$,  instead   of  evaluating  it   separately  in
$\overline{B}$ and $B$, we give the approximation as follows,
\begin{align}
P_{X_{\Sigma}} & =  P_{X_{\Sigma} \vert \overline{B}} \Pr(\overline{B}) + P_{X_{\Sigma} \vert B} \Pr(B) \nonumber \\
& \approx  P_{X_{\Sigma} \vert \overline{B}} \nonumber \\
& =  P_{X_k \vert \overline{B}} * p_{X_k \vert \overline{B}}^{(k-1)*} \nonumber \\
& \approx     P_{X_k} * p_{X_k}^{(k-1)*}
\label{eq:somme}
\end{align}
where   $*$   stands  for   convolution,   and   $(k-1)*$   the  $(k-1)$   times
convolution. When  $\Pr(\overline{B}) \gg \Pr(B)$ (This  hypothesis is realistic
in the case  where $T \gg \Delta t$  and $F \gg \Delta f$, which  is verified in
most       LPWANs       scenarios~\cite{centenaro2015long,      margelis2015low,
  vangelista2015long,           mikhaylov2016analysis,           do2014benefits,
  adelantado2016understanding,  reynders2016range}), the first  approximation is
obviously valid.  The second  approximation comes from  $P_{X_k} =  P_{X_k \vert
  \overline{B}} \Pr(\overline{B}) + P_{X_k  \vert B} \Pr(B) \approx P_{X_k \vert
  \overline{B}}$,  and it's  the  same  with $p_{X_k}$.   We  use $P_{X_k}$  and
$p_{X_k}$  obtained in  the  case  of independent  and  uniform distribution  in
section \ref{sec:1D} and \ref{sec:2D} to evaluate (\ref{eq:somme}).

The reasoning and  the evaluation of $X_{\Sigma}$ in the 1D  case is similar and
thus omitted.

\section{Application}
\label{App:sec}

Let us now illustrate  how our model can be used to  evaluate the performance of
two LPWA technologies.

\subsection{Sigfox{\textsuperscript      \textregistered}}
\subsubsection{2D ``cards tossing'' parameters}
First, we consider  Sigfox{\textsuperscript \textregistered}, an LPWA technology
based  on Ultra  Narrow Band  (UNB)~\cite{goursaud2016random}. The  packet takes
only a  bandwidth $\Delta  f$ around  100\si{Hz}. In doing  so, the  noise power
$\gamma  \Delta  f$  is greatly  reduced  and  the  transmission range  is  thus
increased. In the  physical layer, binary phase-shift keying  (BPSK) is used. In
the   medium    access   control   (MAC)   layer,   random    FDMA   scheme   is
adopted~\cite{do2014benefits}, \ie due  to transmitter oscillator's jitter, it's
not possible  to channelize,  so a  packet is transmitted  at a  randomly chosen
frequency in the available frequency band of 40\si{kHz}.

Sigfox{\textsuperscript \textregistered} also limits  the number of messages per
node to 140  messages per day, which equals to a  message around every 617\si{s}
\cite{reynders2016range}.  $T$ is fixed  to $617\si{s}$, \ie devices transmit as
frequently  as possible.  The  maximal allowed  payload  size per  packet is  12
bytes.  With  the  preamble  and  cyclic  redundancy  check  (CRC)  fields,  the
transmission    duration     $\Delta    t$    of    a     packet    is    around
1.76\si{s}~\cite{goursaud2015dedicated}.  Note  that  this  $\Delta t$  and  $T$
satisfy the  European Telecommunications Standards  Institute (ETSI) requirement
of 1\% duty cycle constraints  in the 868\si{MHz} band~\cite{etsi}.  The ``cards
tossing''  game of  Sigfox{\textsuperscript \textregistered}  falls into  the 2D
case described in \ref{sec:2D} and  its parameters, \ie $\Delta f$, $F$, $\Delta
t$ and $T$ are given in table \ref{tab:lora}.

\subsubsection{Outage and throughput model for Sigfox{\textsuperscript      \textregistered}}
For   now,   there  is   no   report  of   any   power   control  mechanism   in
Sigfox{\textsuperscript   \textregistered},   so   the   model   introduced   in
\ref{sec:outage1}  is   chosen.   Parameters  $\rho_{tm}$,   $\gamma$,  $\zeta$,
$\beta$, $r_{\max}$ are  also listed in table \ref{tab:lora}.  The derivation of
these parameters is as follows.

The  maximum  transmission power  in  868\si{MHz}  is  fixed to  14\si{dBm}  \ie
$\rho_{tm}    \Delta    f   =    14\si{dBm}$~\cite{etsi}.    Thanks   to    UNB,
Sigfox{\textsuperscript  \textregistered} benefits  from a  reduced  noise floor
around -154\si{dBm}, \ie $\gamma \Delta f  = -154\si{dBm}$. This gives us a link
budget around  168\si{dB}. Let's consider  a reception threshold of  8\si{dB}, a
shadow  fading margin  of 10\si{dB}  as  well as  a penetration  loss of  around
15\si{dB}  for  urban  environment.  This  gives  us  a  target  $\SINR$  around
33\si{dB}, \ie $\zeta(\si{dB}) = 33\si{dB}$.  Finally let's consider a path loss
exponent $\beta$ of 3.6 for urban environment. All of these parameters give us a
$r_{\max}$ of 5.2\si{km} for urban scenario.

\begin{table}[ht]
\centering
\caption{Table of Notations}
    \renewcommand{\arraystretch}{1.2}
    \begin{tabular}{|l|c|c|}
    \hline
     & Sigfox{\textsuperscript      \textregistered}  & LoRa{\textsuperscript      \textregistered}  \\ \hline
    Outage and throughput model  &  With path loss   &  Perfect  power\\ 
		 & and fading  &    control \\ \hline
     ``cards tossing'' 	model  & 2D   & 1D   \\ \hline
	 Transmission power & 14 & See section \ref{sec:PCLoRaWAN}  \\ 
	$\rho_{tm}$ $\Delta f $(\si{dBm}) &  &   \\ \hline
	 Noise floor $\gamma \Delta f$(\si{dBm}) & -154 &  -117 \\ \hline
	Target $\SINR \zeta$ (\si{dB}) & 33 &  See table \ref{tab:loraSF} \\ \hline
	 Path loss exponent $\beta$ & 3.6 & 3.6  \\ \hline
	Range $r_{\max}$ (\si{km})& 5.2 & See table \ref{tab:loraSF}  \\ \hline
	Cell form & Single annulus & Multiple annuli \\ \hline
	Payload size (bytes) & 12 & See table \ref{tab:loraSF} \\ \hline
	Application bit rate (\si{bits/s}) & 54.5& See table \ref{tab:loraSF} \\ \hline
   F(\si{kHz}) & 40  &  125 \\ \hline
	  $\Delta f$ & 100\si{Hz}  &  125\si{kHz} \\ \hline
		T(\si{s})  & 617  &  See table \ref{tab:loraSF} \\ \hline
		  $\Delta t$(\si{s})  & 1.76   &  See table \ref{tab:loraSF} \\ \hline
			  Number of channels  & 1  &  3 \\ \hline
    \end{tabular}
\label{tab:lora}
\end{table}

\subsubsection{Simulation}
\begin{figure}[h!]
\centerline{\includegraphics[width=\columnwidth]{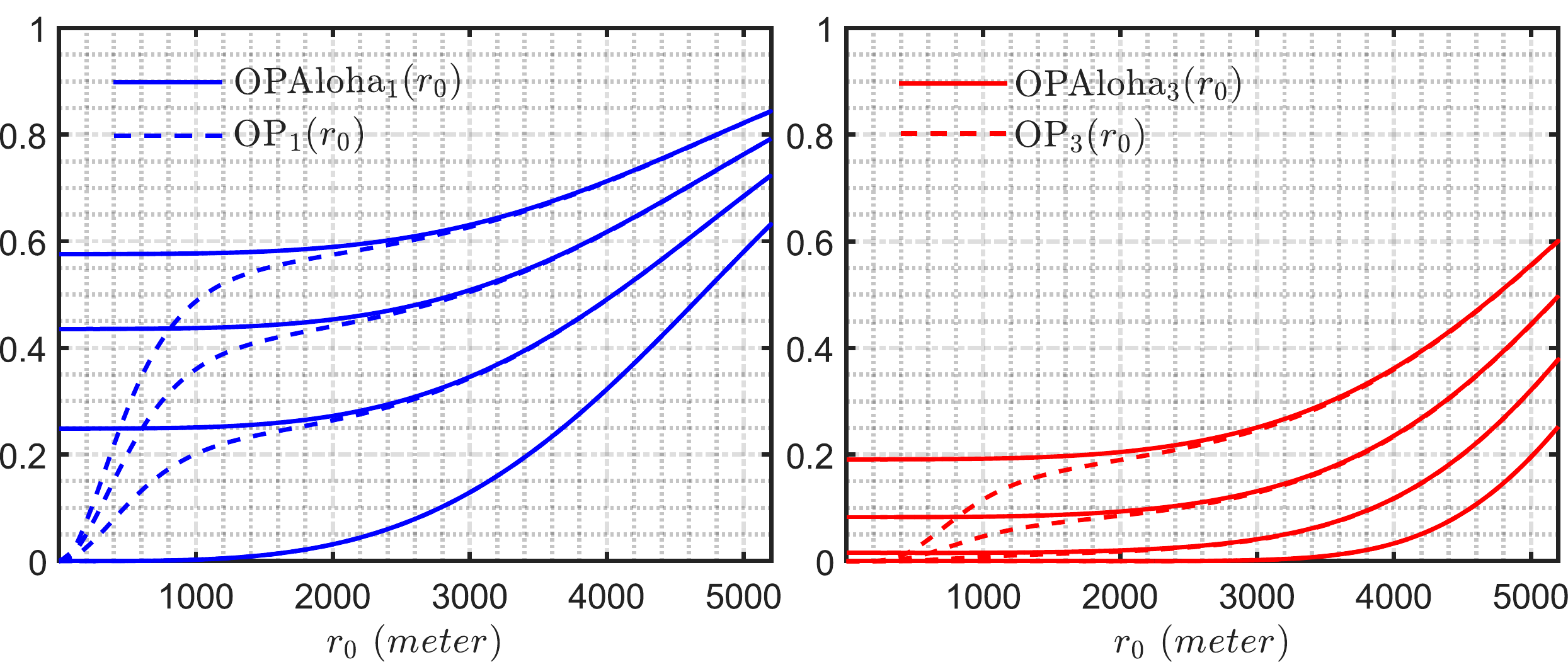}}
\caption{  Outage probability  in  function of  $r_0$  and $N$.  The solid  line
  represents the pure Aloha case, the  dashed line the case with capture effect.
  The five series of curves in each  sub figure represent from top to bottom the
  case where $N = 30000,20000,10000,1$. }
\label{fig:OP12}
\end{figure}

\begin{figure}[h!]
    \centerline{\includegraphics[width=\columnwidth]{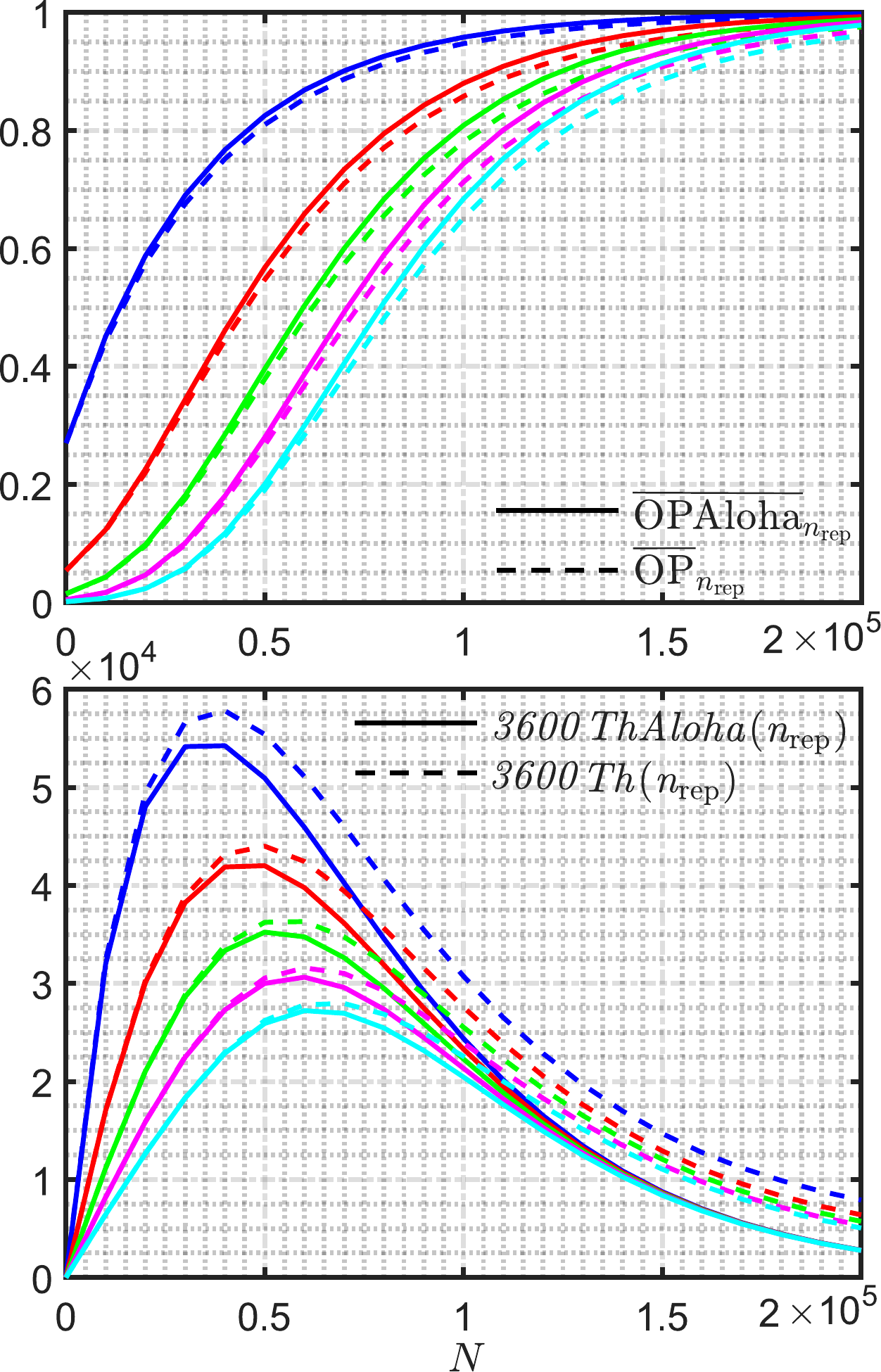}}
    \caption{Global outage  probability and average effective  throughput in one
      hour  for  Sigfox{\textsuperscript  \textregistered}  scenario.  The  five
      colors in each sub figure from top to bottom represent $n_{\mathrm{rep}} =
      1,3,5,7,9$. Solid  line represents  the pure Aloha  case, dashed  line the
      case with capture effect.}
\label{fig:OPSigfox}
\end{figure}

Taking all the  parameters of Sigfox{\textsuperscript \textregistered}, formulae
~\eqref{eq:OPr0rep}--\eqref{eq:OPavg}--\eqref{eq:Throughput}--\eqref{eq:OPAloha}
expressing the first 6 quantities listed in \ref{tab:simulQ} (For the definition
of  $\it{SF}$, see  section \ref{sec:lora})  are simulated  and the  results are
given in figures \ref{fig:OP12} and \ref{fig:OPSigfox}.

\begin{table}[ht]
\centering
\caption{Table of simulated quantities}
    \renewcommand{\arraystretch}{1.1}
    \begin{tabular}{|l|l|}
    \hline
$\OP_{n_{\mathrm{rep}}}(r_0)$ & Outage probability in function of $r_0$ and $n_{\mathrm{rep}}$\\ \hline
$\overline{\OP}_{n_{\mathrm{rep}}}$ &  Global outage probability  \\ 
&   averaged over $r_0$ or $SF$ \\ \hline  
$\it{Th}(n_{\mathrm{rep}})$& Average effective throughput in function of $n_{\mathrm{rep}}$               \\ \hline 
$\OPAloha_{n_{\mathrm{rep}}}(r_0)$&   Outage probability in pure Aloha scenario               \\  
&  in function of $r_0$ and $n_{\mathrm{rep}}$         \\ \hline 
$\overline{\OPAloha}_{n_{\mathrm{rep}}}$&     Global   outage probability in pure           \\ 
&Aloha scenario averaged  over $r_0$ or $SF$
\\ \hline 
$\it{ThAloha}(n_{\mathrm{rep}})$&   Average effective throughput in function of $n_{\mathrm{rep}} $            \\ 
& in pure Aloha scenario      \\ \hline 
$\OP_{n_{\mathrm{rep}}}(SF)$&   Outage probability in function of $SF$ and   $n_{\mathrm{rep}}$             \\ \hline 
$\it{Th}_{SF}(n_{\mathrm{rep}})$&  Average  effective  throughput   of a certain $SF$            \\ 
&    in function of $n_{\mathrm{rep}}$          \\ \hline 
    \end{tabular}
\label{tab:simulQ}
\end{table}

Figure \ref{fig:OP12}  shows that capture  effect represented by  the difference
between the  solid and dashed lines  decreases with the  distance $r_0$, because
naturally the  devices nearer to the  base station have more  chances to benefit
from the capture effect. Repetitions do reduce the outage probability.

Figure \ref{fig:OPSigfox} shows  that $\overline{\OP}_{n_{\mathrm{rep}}}$ is not
greatly reduced in the capture case than in the pure Aloha case; the improvement
in  $\it{Th}(n_{\mathrm{rep}})$ increases  with $N$,  as more  collisions happen
with  higher $N$, thus  amplifying the  capture effect,  but the  high collision
regime is  not the optimal zone  for the low-power devices  to function.  Device
density  $p_R(r)$  increases with  distance  $r_0$  and  further devices  barely
benefit from  the capture effect, so devices  at the cell edge  are probably the
bottleneck  of the network  performance, \ie  it's them  that stops  the network
performance from  getting better. One  can also observe that  repetitions reduce
the  global outage  probability but  also result  in lower  effective throughput
because of the introduced redundancy.   Our abacuses permits to find the optimal
$n_{\mathrm{rep}}$ in function of the target outage probability, $N$, throughput
and energy cost.

\subsection{LoRaWAN{\textsuperscript      \textregistered}}
\label{sec:lora}
\subsubsection{1D ``cards tossing'' parameters }

LoRaWAN{\textsuperscript  \textregistered} is another  LPWA technology  based on
spectrum spreading \cite{lora}. The spreading  factor is denoted by $SF$ and can
vary from 6 to 12. Every packets  are spread in the available bandwidth $F$, \ie
$\Delta f = F$. In Europe 3  default channels are used, each with a bandwidth of
$125\si{kHz}$~\cite{lora}.  Different  $SF$ result  in different bit  rate.  The
smaller the  $SF$, the  higher the  bit rate. Different  payload sizes  are also
specified for different $SF$. $\Delta t_{SF}$ can thus be calculated, the detail
can be found in~\cite{mikhaylov2016analysis,  lora2}.  $T_{SF}$ is fixed to $100
\Delta t_{SF}$ according to the  ETSI duty cycle constraint of $1\%$~\cite{etsi}
in  the 868\si{MHz}  band.   Transmissions from  different  $SF$ are  considered
orthogonal  and do  not interfere  with  each other~\cite{mikhaylov2016analysis,
  reynders2016range,    lora2},    so   the    ``cards    tossing''   game    of
LoRaWAN{\textsuperscript \textregistered}  can be  seen as seven  orthogonal and
parallel  1D  games  described   in  section  \ref{sec:1D},  each  having  three
orthogonal channels  of bandwidth $F  = 125\si{kHz}$, different  $\Delta t_{SF}$
and  $T_{SF}$.    The  corresponding  game   parameters  are  listed   in  table
\ref{tab:lora} and \ref{tab:loraSF} ~\cite{mikhaylov2016analysis,lora2}.

\subsubsection{Multiple annuli cell structure}
\label{sec:multiAnnuli}
The  greater the  $SF$, the  lower  the sensitivity  of transmission  associated
~\cite{lora2} . This results in  smaller $\zeta_{SF}$ required for greater $SF$.
Transmission with  greater $SF$ can thus  reach further. $r_{SF}$  is defined as
the maximum distance  that a transmission with a certain  $SF$ can barely reach,
with only path loss  considered, \ie $\frac{\rho_{tm} r_{SF}^{-\beta}}{\gamma} =
\SNR = \zeta_{SF}$, where $\rho_{tm}  \Delta f$ takes the maxmimum allowed power
$14\si{dBm}$~\cite{etsi}.  The noise power  is around $-117\si{dBm}$.  Let's add
a  shadow margin  of 10  \si{dB} as  well  as a  penetration loss  of around  15
\si{dB},  to   give  us   the  $\zeta_{SF}$  for   different  $SF$,   see  table
\ref{tab:loraSF}. The  path loss  exponent $n$ is  fixed to  the same 3.6  as in
Sigfox{\textsuperscript  \textregistered} scenario.   With  the maximal  allowed
transmission power  $\rho_t \Delta f$  of 14 \si{dBm}, the  communication ranges
$r_{SF}$ in terms of path loss for different $SF$ are thus calculated and listed
in table \ref{tab:loraSF}. These are the ranges with which the reception $\SNR =
\frac{\rho_t r_{SF}^{-n}}{\gamma}$ barely satisfies $\zeta_{SF}$, in presence of
only  path   loss  and  without   considering  fading  and   interferences,  \ie
$\zeta_{SF}^{-1}-\SNR^{-1}=0$.

Further devices should use greater $SF$ to simply reach out to the base station,
while nearer  devices can benefit  from higher bit  rate of smaller  $SF$. Let's
consider a ideally  pre-configured network where all the  devices located in the
annulus defined by $r_{SF}$ and $r_{{SF}-1}$ take spreading factor $SF$ (For $SF
\, 6$, it's  the annulus between $r_6$ and  $r_c$) so that they make  use of the
smallest possible $SF$ and thus the highest possible bit rate while guaranteeing
the   communication   range   at  the   same   time~\cite{mikhaylov2016analysis,
  adelantado2016understanding}.   The  probability  of  a node  falling  into  a
certain    annulus    denoted   by    $p_{SF}$    is    proportional   to    its
surface~\cite{adelantado2016understanding}.   The  number  of devices  taking  a
certain $SF$ is  thus just $N p_{SF}$.  $\zeta_{SF}$,  $r_{SF}$ and $p_{SF}$ are
listed in table \ref{tab:loraSF}.

LoRaWAN{\textsuperscript    \textregistered}   network    support   over-the-air
activation of  the node which  requires the node  to open 2  successive downlink
windows after a uplink transmission, in  order to receive the MAC layer commands
from  the network~\cite{lora}.  Also,  LoRaWAN{\textsuperscript \textregistered}
network infrastructure  can manage  the $SF$ and  date rate  by means of  an ADR
(Adaptive  Data Rate)  scheme, which  also necessitates  the node  to  listen to
gateway  downlink   transmissions.  Note  that   the  downlink  and   uplink  in
LoRaWAN{\textsuperscript \textregistered} share  the same channels, so collision
phenomenon may be aggravated by  the use of downlink. The pre-configured network
that we consider  is the scenario without nodes listening  to gateway, \ie there
is only uplink transmissions. Each node  is set to an appropriate $SF$ according
to its distance from the gateway.

\begin{table}[ht]
\centering
\caption{Parameters of LoRaWAN{\textsuperscript      \textregistered}}
    \renewcommand{\arraystretch}{1.2}
		{\setlength{\tabcolsep}{0.1cm}} 
     \begin{tabular}{|l|c|c|c|c|c|c|c|c|}
     \hline
     SF & Sensitivity  & $\zeta_{SF}$& Range & $p_{SF}$ &Payload&$\Delta t_{SF}$ & Bit rate &$T_{SF}$\\ 
 		 &(\si{dBm}) &(\si{dB}) &$r_{SF}$(\si{km})  & & (bytes)& (\si{s}) &  (\si{Kb/s})& (\si{s})\\ \hline
     6 &  -121  & 21 &1.13 & 0.13 & 242 &0.233&8.309 & 23.3\\ \hline
     7 & -124 & 18 &1.37 & 0.06 & 242 & 0.400&4.840 &40.0\\ \hline
    8& -127 & 15 &1.67&  0.09 & 242 &0.707&2.738&70.7\\ \hline
 	9& -130& 12 &2.02 &  0.13 & 115 & 0.677& 1.359 &67.7\\ \hline
 	10& -133 & 9&2.45 &  0.19 & 51&0.698& 0.585&69.8\\ \hline
 		11& -135 & 7 &2.78 & 0.17 & 51 &1.561& 0.261&156.1\\ \hline
 		12& -137 & 5 &3.16 &  0.23 &51 &2.793& 0.146&279.3\\ \hline
 	\end{tabular}
\label{tab:loraSF}
\end{table}

\subsubsection{Outage  and  throughput  model  with  power  control  scheme  for
  LoRaWAN{\textsuperscript \textregistered}}
\label{sec:PCLoRaWAN}
To improve  the fairness of devices located  in the same annulus  and having the
same $SF$, we consider the  following ideal power allocation strategy : Allocate
$\rho_{tm} \Delta f = 14\si{dBm}$ to  the devices with distance $r_{SF}$ to make
sure they get covered; Make sure  that the reception power density attenuated by
path loss  of all devices in the  same annulus, denoted by  $\rho$, is identical
and  equal   to  that  of   devices  with  distance  $r_{SF}$,   \ie  $\rho_{tm}
r_{SF}^{-\beta}$. Fading  effect is neglected  in this case. This  setting falls
into   the  perfect   power   control  paradigm   introduced   in  sub   section
\ref{sec:powerC}.  $\zeta_{SF}^{-1} - \SNR^{-1}  = 0$  according to  the section
\ref{sec:multiAnnuli}. By shrinking all  the annuli, our power allocation scheme
can  in fact  result in  non  zero $\zeta_{SF}^{-1}  - \SNR^{-1}$.  With cdf  of
$X_{\Sigma}$ already  given in section~\ref{Proba:sec},  there is no  problem in
evaluating this adaptation.

In  our scenario,  the  non fairness  between  devices with  the  same $SF$  but
different distance $r_0$ is removed, \ie $\OP_{n_{\mathrm{rep}}}$ doesn't depend
on $r_0$ any  more, but non fairness exists between different  $SF$ as $SF$ with
greater  $p_{SF}$  have  greater   device  number  $Np_{SF}$  and  thus  greater
$\OP_{n_{\mathrm{rep}}}(SF)$         expressed         as         $\displaystyle
\OP_{n_{\mathrm{rep}}}(SF)   =   \left[\Pr\left[X_{\sum_{k=1}^{N_{SF}}  }   \geq
    0\right]\right]^{n_{\mathrm{rep}}}$. $\it{Th}(n_{\mathrm{rep}}) = \sum_{SF =
  6}^{12} \it{Th}_{SF}(n_{\mathrm{rep}})$ should be recast as follows,
\begin{equation}
\it{Th}(n_{\mathrm{rep}})  = \sum_{SF = 6}^{12}  \frac{3 N p_{SF}(1-\OP_{n_{\mathrm{rep}}}(SF))}{T_{SF} \, n_{\mathrm{rep}}}
\end{equation}
where  the factor  three comes  from the  three available  channels.  The outage
probability       averaged      over       $SF$       is      expressed       as
$\overline{\OP}_{n_{\mathrm{rep}}}       =        \sum_{SF       =       6}^{12}
\OP_{n_{\mathrm{rep}}}(SF)p_{SF}$.  Quantities  to be  simulated  are listed  in
table    \ref{tab:simulQ}.    Note   that    in    the   scenario    considered,
$\overline{\OP}_{n_{\mathrm{rep}}}$                coincides                with
$\overline{\OPAloha}_{n_{\mathrm{rep}}}$,  and  $\it{Th}(n_{\mathrm{rep}})$ with
$\it{ThAloha}(n_{\mathrm{rep}})$ as there is no tolerance of overlapping.

\subsubsection{Simulation}
\begin{figure}[h!]
    \centerline{\includegraphics[width=\columnwidth]{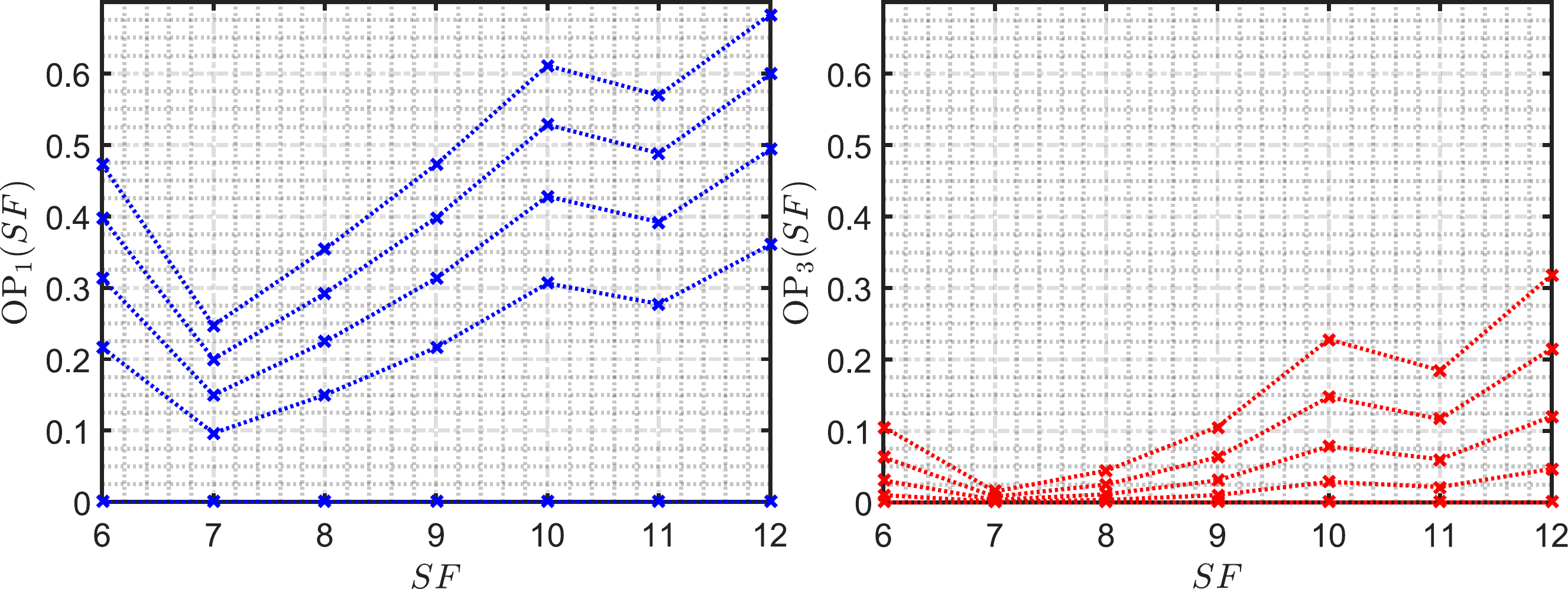}}
    \caption{Outage probability  of LoRaWAN{\textsuperscript \textregistered} in
      function of $N$ and $SF$. From top  to bottom, the five curves in each sub
      figure represent $N  = 250, 200, 150,  100, 1$. Sub figure on  the left is
      the case  where $n_{\mathrm{rep}} =  1$, on the right  $n_{\mathrm{rep}} =
      3$. }
\label{fig:LoraOP1}
\end{figure}

\begin{figure}[h!]
\centerline{\includegraphics[width=\columnwidth]{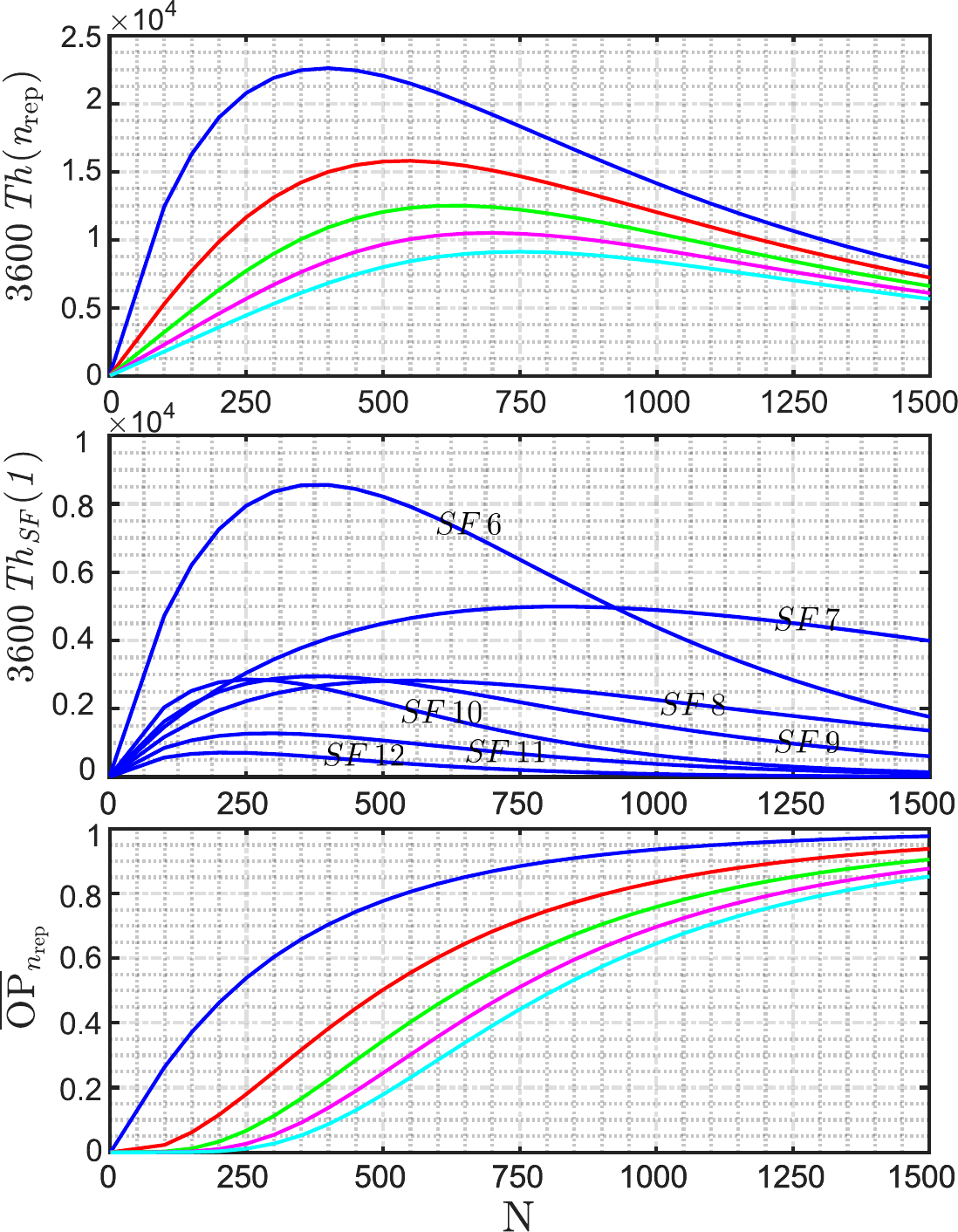}}
\caption{The figure  in the top  represents average effective throughput  in one
  hour.  The  five  different colors  from  top  to  bottom represent  the  case
  $n_{\mathrm{rep}}  = 1,  3, 5,  7,  9$. The  figure in  the middle  represents
  average   effective  throughput   in  one   hour  of   different   $SF$,  with
  $n_{\mathrm{rep}}=1$. The numbers on the curves stand for the $SF$. The figure
  in the  bottom represents the global  outage probability. The  colors have the
  same meaning as described earlier.  }
\label{fig:OPThLoRA}
\end{figure}

The simulation results in the LoRaWAN{\textsuperscript \textregistered} case are
given    in   figures    \ref{fig:LoraOP1}   and    \ref{fig:OPThLoRA}.   Figure
\ref{fig:LoraOP1} shows the non fairness  in terms of outage probability between
different $SF$, which  is directly related to $p_{SF}$,  which again is dictated
by non uniformity of device density in function with $r_0$.

Several  observations  can  be   made  from  figure  \ref{fig:OPThLoRA}.  First,
repetition  mechanism reduces overall  outage probability  but also  the average
effective throughput  as redundancy is  introduced. It increases energy  cost as
well.  Second,  the  differences  between  $\it{Th}_{SF}(n_{\mathrm{rep}})$  are
multiple,   dictated    by   $p_{SF}$   and    $T_{SF}$.   $p_{SF}$   determines
$OP_{n_{\mathrm{rep}}}(SF)$,  while  $T_{SF}$ determines  the  speed of  message
sending. For example, $SF \, 6$ is the  fastest of all $SF$, but as $p_6 > p_7$,
the curve  of $SF \, 6$  reaches its maximum  much earlier than $SF  \,7$, which
limits  its performance. $SF  \, 12$  has the  worst performance  as it  has the
longest $T_{SF}$ and biggest $p_{SF}$,  which again confirms that the devices at
the cell edge are probably the bottleneck of the network performance.

At  last, if  we  try to  compare  Sigfox{\textsuperscript \textregistered}  and
LoRaWAN{\textsuperscript   \textregistered},  we  can   see  that   even  though
Sigfox{\textsuperscript \textregistered}  can support more devices  but in terms
of   throughput,   it's   of   the  same   order   as   LoRaWAN{\textsuperscript
  \textregistered}.  Note  that the  payload  sizes in  LoRaWAN{\textsuperscript
  \textregistered} are much more  important than that of Sigfox{\textsuperscript
  \textregistered}  (see table  \ref{tab:lora} and  \ref{tab:loraSF}).  It seems
that Sigfox{\textsuperscript \textregistered}  is more suitable for applications
with  a lot of  devices having  smaller traffic,  while LoRaWAN{\textsuperscript
  \textregistered} can support applications with more important traffic but less
devices.


\section{Conclusion and perspectives }
\label{conclu:sec}

In  this article,  we  provide a  high-level  flexible model  whose interest  is
illustrated by the performance evaluation  of two LPWA technologies. To the best
of our knowledge,  this is the first model  which considers joint time-frequency
interference. Note that our paradigm can be adapted to other systems to evaluate
the   relationship  between   number  of   devices,  repetition   times,  outage
probability,  throughput and  energy cost.  Capture  effect is  also taken  into
account  in  our  model  so  further   questions  such  as  how  to  amplify  it
intelligently  can be investigated  in the  future.  We  believe that  our model
provides a useful dimensioning tool for the future IoT scenarios.

Our model can be completed in the mathematical level. First, when the hypotheses
$T \gg  \Delta t$ and $ F\gg  \Delta f$ are not  satisfied, algorithmic approach
seems to be more adapted due  to the difficulty in the probability evaluation on
the border area. Second, the proportion between $\Delta t$ and $\Delta f$ has an
influence on the  probability distributions of $X_k$ and  $X_\Sigma$, thus comes
the  question  of the  best  strategy of  proportioning  the  time duration  and
frequency  occupancy  of  the  information  packet, in  order  to  minimize  the
overlapping.   At last, it's  possible to  formulate a  more general  problem of
finding the best  strategy to use a 2D time-frequency  resource, always in terms
of  minimization  of  overlapping   phenomenon.  We  have  the  options  between
orthogonal  division of  the frequency  band, division  into  multiple partially
overlapping bands (POC), and random FDMA if we go to the extreme.

In  both the  LPWANs scenarios  considered,  cell edge  devices seem  to be  the
bottleneck  of  the global  network  performance.  A  possible solution  is  the
densification of  the infrastructure, knowing  that IoT devices  can communicate
with multiple base stations \ie  multiple reception or macro-diversity.  A study
on  the  $k$-coverage  of devices  is  given  in  \cite{hal}. By  combining  the
$k$-coverage model  for LPWANs and our  ``cards tossing'' model,  we can jointly
design  the infrastructure  deployment and  MAC layer  of devices,  in  order to
improve the network performance while limiting the cost.

\bibliography{ICC}

\begin{thebibliography}{10}

\bibitem{centenaro2015long}
Marco Centenaro, Lorenzo Vangelista, Andrea Zanella, and Michele Zorzi.
\newblock Long-range communications in unlicensed bands: The rising stars in
  the {I}o{T} and smart city scenarios.
\newblock {\em arXiv preprint arXiv:1510.00620}, 2015.

\bibitem{margelis2015low}
George Margelis, Robert Piechocki, Dritan Kaleshi, and Paul Thomas.
\newblock Low throughput networks for the {I}o{T}: Lessons learned from
  industrial implementations.
\newblock In {\em Internet of Things (WF-IoT), 2015 IEEE 2nd World Forum on},
  pages 181--186. IEEE, 2015.

\bibitem{vangelista2015long}
Lorenzo Vangelista, Andrea Zanella, and Michele Zorzi.
\newblock Long-range {I}o{T} technologies: The dawn of {L}o{R}a\texttrademark.
\newblock In {\em Future Access Enablers of Ubiquitous and Intelligent
  Infrastructures}, pages 51--58. Springer, 2015.

\bibitem{Abr09}
N.~Abramson.
\newblock The {A\sc{LOHA}N}et -- surfing for wireless data.
\newblock {\em IEEE Comm. Magazine}, 47(12):21--25, 2009.

\bibitem{TanWet11}
A.~S. Tanenbaum and D.~J. Wetherall.
\newblock {\em Computer Networks}.
\newblock Prentice Hall, Boston, 5th edition, 2011.

\bibitem{ware2001modelling}
Christopher Ware, Joe Chicharo, and Tadeusz Wysocki.
\newblock Modelling of capture behaviour in {ieee} 802.11 radio modems.
\newblock In {\em IEEE Int. Conf. on Telecomm.}, 2001.

\bibitem{arnbak1987capacity}
JENSC Arnbak and Wim Van~Blitterswijk.
\newblock Capacity of slotted {ALOHA} in {R}ayleigh-fading channels.
\newblock {\em IEEE Journal on Selected Areas in Comm.}, 5(2):261--269, 1987.

\bibitem{cheun1998joint}
Kyungwhoon Cheun and Sunyoung Kim.
\newblock Joint delay-power capture in spread-spectrum packet radio networks.
\newblock {\em IEEE Transactions on Comm.}, 46(4):450--453, 1998.

\bibitem{lee2007experimental}
Jeongkeun Lee, Wonho Kim, Sung-Ju Lee, Daehyung Jo, Jiho Ryu, Taekyoung Kwon,
  and Yanghee Choi.
\newblock An experimental study on the capture effect in 802.11 a networks.
\newblock In {\em Proceedings of the second ACM international workshop on
  Wireless network testbeds, experimental evaluation and characterization},
  pages 19--26, 2007.

\bibitem{mishra2006partially}
Arunesh Mishra, Vivek Shrivastava, Suman Banerjee, and William Arbaugh.
\newblock Partially overlapped channels not considered harmful.
\newblock In {\em ACM SIGMETRICS Performance Evaluation Review}, volume~34,
  pages 63--74, 2006.

\bibitem{villegas2007effect}
Eduard~Garcia Villegas, Elena Lopez-Aguilera, Rafael Vidal, and Josep
  Paradells.
\newblock Effect of adjacent-channel interference in {IEEE} 802.11 {WLAN}s.
\newblock In {\em 2007 2nd international conf. on cognitive radio oriented
  wireless networks and comm.}, pages 118--125. IEEE, 2007.

\bibitem{xing2009multi}
Guoliang Xing, Mo~Sha, Jun Huang, Gang Zhou, Xiaorui Wang, and Shucheng Liu.
\newblock Multi-channel interference measurement and modeling in low-power
  wireless networks.
\newblock In {\em Real-Time Systems Symposium, 2009, RTSS 2009. 30th IEEE},
  pages 248--257. IEEE, 2009.

\bibitem{mikhaylov2016analysis}
Konstantin Mikhaylov, Juha Pet{\"a}j{\"a}j{\"a}rvi, and Tuomo Haenninen.
\newblock Analysis of capacity and scalability of the {L}o{R}a {L}ow {P}ower
  {W}ide {A}rea {N}etwork {T}echnology.
\newblock In {\em European Wireless 2016; 22th European Wireless Conference;
  Proceedings of}, pages 1--6. VDE VERLAG GmbH, 2016.

\bibitem{goursaud2016random}
Claire Goursaud and Yuqi Mo.
\newblock Random unslotted time-frequency aloha: Theory and application to iot
  unb networks.
\newblock In {\em Telecommunications (ICT), 2016 23rd International Conference
  on}, pages 1--5. IEEE, 2016.

\bibitem{adelantado2016understanding}
Ferran Adelantado, Xavier Vilajosana, Pere Tuset-Peiro, Borja Martinez, and
  Joan Melia.
\newblock Understanding the limits of {L}o{R}a{WAN}.
\newblock {\em arXiv preprint arXiv:1607.08011}, 2016.

\bibitem{reynders2016range}
Brecht Reynders, Wannes Meert, and Sofie Pollin.
\newblock Range and coexistence analysis of long range unlicensed
  communication.
\newblock In {\em Telecommunications (ICT), 2016 23rd International Conference
  on}, pages 1--6. IEEE, 2016.

\bibitem{li2015internet}
Shancang Li, Li~Da~Xu, and Shanshan Zhao.
\newblock The internet of things: a survey.
\newblock {\em Information Systems Frontiers}, 17(2):243--259, 2015.

\bibitem{aljuaid2010investigating}
Muhammad Aljuaid and Halim Yanikomeroglu.
\newblock Investigating the gaussian convergence of the distribution of the
  aggregate interference power in large wireless networks.
\newblock {\em {IEEE} Trans. On Vehicular Tech.}, 59(9):4418--4424, 2010.

\bibitem{do2014benefits}
Minh-Tien Do, Claire Goursaud, and Jean-Marie Gorce.
\newblock On the benefits of random {fdma} schemes in ultra narrow band
  networks.
\newblock In {\em Modeling and Optimization in Mobile, Ad Hoc, and Wireless
  Networks (WiOpt), 2014 12th Int. Symposium on}, pages 672--677. IEEE, 2014.

\bibitem{goursaud2015dedicated}
Claire Goursaud and Jean-Marie Gorce.
\newblock Dedicated networks for {I}o{T}: {PHY}/{MAC} state of the art and
  challenges.
\newblock {\em EAI endorsed transactions on Internet of Things}, 2015.

\bibitem{etsi}
{ETSI}.
\newblock
  \url{http://www.etsi.org/deliver/etsi_tr/103000_103099/103055/01.01.01_60/tr_103055v010101p.pdf}.
\newblock Accessed: 2016-10-24.

\bibitem{lora}
{LoRaWAN\texttrademark~Specification}.
\newblock
  \url{https://www.lora-alliance.org/portals/0/documents/whitepapers/LoRaWAN101.pdf}.
\newblock Accessed: 2016-10-24.

\bibitem{lora2}
{LoRaWAN\texttrademark~Tranceiver Specification}.
\newblock \url{http://www.semtech.com/images/datasheet/sx1272.pdf}.
\newblock Accessed: 2016-10-24.

\bibitem{hal}
Zhuocheng Li, Tuong-Bach Nguyen, Quentin Lampin, Isabelle Sivignon, and Steeve
  Zozor.
\newblock Ensuring k-coverage in low-power wide area networks for internet of
  things.
\newblock \url{https://hal.archives-ouvertes.fr/hal-01353801}.

\end{thebibliography}
\bibliographystyle{unsrt}
\end{document}